\documentstyle[11pt,appb1]{article} 
\def\be{\begin{equation}}
\def\ee{\end{equation}}
\def\grav{\tilde{G}}
\def\ov{\overline}
\def\slash#1{\setbox0=\hbox{$#1$}#1\hskip-\wd0\dimen0=5pt\advance
       \dimen0 by-\ht0\advance\dimen0 by\dp0\lower0.5\dimen0\hbox
         to\wd0{\hss\sl/\/\hss}}
\begin{document}
\title{LOW-ENERGY GRAVITINO INTERACTIONS\thanks{Presented at 
the International Europhysics Conference on High Energy Physics,
Jerusalem, Israel, 19-26 August 1997 and at the XXI
International School of Theoretical Physics
``Recent Progress in Theory and
Phenomenology of Fundamental Interactions'', 
Ustron, Poland, 19-24 September 1997.}
}
\author{Ferruccio Feruglio
\address{Theory Division, CERN, Switzerland\\ 
on leave from University of Padova, Italy}
}
\maketitle
\begin{abstract}
I discuss the low-energy limit of several processes involving only ordinary 
particles and gravitinos. Astrophysical and laboratory applications are briefly
addressed.
\end{abstract}
\vskip 0.2cm

It is well known that the solution of the hierarchy problem in 
supersymmetric extensions of the Standard Model (SM) requires
a mass splitting $\Delta m \sim 1$ TeV between ordinary particles and 
their superpartners.
This requirement, however, leaves largely undetermined the 
supersymmetry-breaking scale $\sqrt{F}$, or, equivalenty, the gravitino mass 
$m_{3/2}=F/(\sqrt{3} M_{P})$, $M_P$ being the Planck mass.
The ratio $\Delta m^2/F$ is given by the coupling of 
the goldstino to the matter sector under consideration. If this coupling is 
gravitational, of order $m_{3/2}/M_{P}$, then $\Delta m$ is of order
$m_{3/2}$ and supersymmetry breaking takes place at the
intermediate scale $\sqrt{F}\sim 10^{10}$ GeV. On the other hand,
if the goldstino coupling to matter is of order 1, then $\sqrt{F}$
is comparable to the mass splitting $\Delta m$, and the gravitino 
becomes superlight, with a mass of about $10^{-5}$ eV. In the absence of
a theory of supersymmetry breaking, $F$ should be treated as a free parameter.

If the gravitino is superlight, then 
one expects a substantially different phenomenology
from that characterizing the Minimal Supersymmetric Standard Model (MSSM).
In this case, only the $\pm 3/2$ gravitino helicity states 
can be safely omitted from the low-energy effective theory,
when gravitational interactions are neglected.
The $\pm 1/2$ helicity states, essentially described by the
goldstino field, should instead be accounted for at low energy,
because of their non-negligible coupling to matter.
The lightest supersymmetric particle is the gravitino and 
peculiar experimental signatures can arise from the decay of
the next-to-lightest supersymmetric particle into its ordinary
partner plus a gravitino \cite{amb}.

Moreover, even when all superymmetric particles of the MSSM are above the 
production threshold, 
interesting signals could come from those processes
where only ordinary particles and gravitinos occur. 
As soon as the typical energy of the process is larger than $m_{3/2}$, 
a condition always fulfilled in the applications discussed below,
one can approximate the physical amplitudes by replacing external
gravitinos with goldstinos, as specified by the equivalence theorem 
\cite{equ}. 
If the masses of the ordinary particles 
involved are negligible with respect to the energy of interest, 
these processes are controlled by just one dimensionful parameter, 
the supersymmetry-breaking scale $\sqrt{F}$, entering the amplitudes
in the combination $(\tilde{G}/\sqrt{2}F)$, $\tilde{G}$ denoting
the goldstino wave function. 
   
This class of processes includes $\gamma\gamma\to\tilde{G}\tilde{G}$,   
$e^+e^-\to\tilde{G}\tilde{G}$, which may influence primordial 
nucleosynthesis, stellar cooling and supernovae 
explosion \cite{bfz1,luty}. 
Of direct interest for LEP2 and for the future
linear colliders is the reaction $e^+e^-\to\tilde{G}\tilde{G}\gamma$.
Partonic reactions such as $q{\bar q}\to\tilde{G}\tilde{G}\gamma$,
$q{\bar q}\to\tilde{G}\tilde{G}g$ and $q g\to\tilde{G}\tilde{G} q$
can be indirectly probed at the Tevatron collider or in future hadron
facilities. In the absence of experimental signals, one can use
these processes to set absolute limits on the gravitino mass.
At variance with other bounds on $m_{3/2}$ discussed in the literature
\cite{bou}, these limits have the advantage of not depending on detailed
assumptions about the spectrum of supersymmetric particles.
Finally, the study of these processes
can reveal unexpected features of the low-energy theory, which
were overlooked in the standard approach to goldstino low-energy
interactions.

The natural tools to analyse the above processes are the so-called
low-energy theorems \cite{fri}.
According to these, 
the low-energy amplitude for the scattering of a goldstino on a given target 
is controlled by the energy--momentum tensor 
$T_{\mu\nu}$ of the target.
To evaluate the physical amplitudes, it is more practical to make use
of an effective Lagrangian, containing the goldstino field
and the matter fields involved in the reactions, and providing a
non-linear realization of the supersymmetry algebra \cite{nli}.
For instance, in the non-linear construction of \cite{wes},
the goldstino field $\tilde{G}$ and the generic matter field 
$\varphi$ are incorporated into the following superfields:
\be
\label{bigl}
\Lambda_{\alpha} 
\equiv \exp (\theta Q + \ov{\theta} \ov{Q} ) \, 
\grav_{\alpha} 
=  {\grav_{\alpha}\over \sqrt{2} F} + \theta_{\alpha} +
{i \over \sqrt{2} F} ( \grav \sigma^{\mu} \ov{\theta} 
- \theta \sigma^{\mu} \ov{\grav} ) \partial_{\mu} 
{\grav_{\alpha}\over \sqrt{2} F} + \ldots  \, ,
\ee
\be
\Phi  \equiv 
\exp (\theta Q + \ov{\theta} \ov{Q} ) \, \varphi 
=  \varphi + {i \over \sqrt{2} F} ( \grav \sigma^{\mu} 
\ov{\theta} - \theta \sigma^{\mu} \ov{\grav} ) \partial_{\mu} 
\varphi + \ldots  \, .
\label{mat}
\ee
The goldstino--matter system is described by the
supersymmetric Lagrangian:
\be
\int d^2\theta d^2{\bar\theta} 
\Lambda^2 {\bar\Lambda}^2 
\left[2 F^2+
{\cal L}(\Phi,\partial\Phi)
\right]~~~,
\label{inv1}
\ee 
where ${\cal L}(\varphi,\partial\varphi)$ is the ordinary Lagrangian for the 
matter system.
This non-linear realization automatically reproduces the results
of the low-energy theorems, in particular the expected 
goldstino coupling to the energy--momentum tensor $T_{\mu\nu}$ associated to 
$\varphi$. 

An alternative approach consists in constructing
a low-energy Lagrangian, starting from a general 
supersymmetric theory defined, up to terms with more than two derivatives,
in terms of a K\"ahler potential, a superpotential and
a set of gauge kinetic functions. The effective theory can be obtained
by integrating out, in the low-energy limit, the heavy superpartners
\cite{bfz1}.

When applied to the process $\gamma\gamma\to\tilde{G}\tilde{G}$,
the two procedures yield the same result. The only independent,
non-vanishing, helicity amplitude for the process is:
\be
a(1,-1,1/2,-1/2)=8\sin\theta\cos^2{\theta\over 2}~{E^4\over F^2}~~~,
\label{aph} 
\ee
where (1, $-$1) and  (1/2, $-$1/2) are the helicities of the incoming 
and outgoing particles, respectively; $E$ and $\theta$ are the
goldstino energy and scattering angle in the centre-of-mass frame.
The total cross section is $s^3/(640\pi F^4)$.

In earlier cosmological and astrophysical applications,
with a typical energy range from about 1 keV to 100 MeV,
a cross-section scaling as $\Delta m^2 s^2/F^4$ was assumed,
giving rise to a lower bound on $m_{3/2}$ close to
$10^{-6}$ eV. When the correct energy dependence is taken into account,
this bound is reduced by at least a factor 10,
and becomes uninteresting compared to those obtainable 
at colliders.

When considering $e^+e^-\to\tilde{G}\tilde{G}$, in the limit of massless
electron, one has to face an unexpected
result \cite{bfz2}. On the one hand, by integrating out the heavy
selectron fields, one finds the following helicity amplitude:
\be
a(1/2,-1/2,1/2,-1/2)={4 (1+\cos\theta)^2 E^4\over F^2}~~~,
\label{ae1}
\ee
all other non-vanishing amplitudes being related to this one.
On the other hand, by using the non-linear realization of \cite{wes},
one obtains:
\be
a(1/2,-1/2,1/2,-1/2)={4\sin^2\theta E^4\over F^2}~~~.
\label{ae2}
\ee
The amplitudes of eqs. (\ref{ae1}) and (\ref{ae2}) scale in the same way 
with the energy, but have a different
angular dependence. We should conclude that the low-energy theorems
of ref. \cite{fri} do not apply to the case of a massless fermion. 
A particularly disturbing aspect is that the non-linear realization of eq. 
(\ref{inv1}) is supposed to provide
the most general parametrization of the amplitude in question,
independently of any considerations about the low-energy theorems.
In the case at hand, the Lagrangian of eq. (\ref{inv1}) reads:
\be
{\cal L}_e=
\int d^2\theta d^2{\bar\theta} 
\Lambda^2 {\bar\Lambda}^2 
\left[2 F^2+
i E\sigma^\mu\partial_\mu {\bar E}+i E^c\sigma^\mu\partial_\mu {\bar E^c}
\right]~~~,
\label{inv1bis}
\ee
where $E$ and $E^c$ are the superfields associated to the two Weyl
spinors $e$ and $e^c$ describing the electron, according to eq. 
(\ref{mat}). 
The solution to this puzzle \cite{bfz2} is provided by the existence
of an independent supersymmetric
invariant that has been neglected up to now in the literature:
\be
\delta{\cal L}_e=\int d^2\theta d^2{\bar\theta}
(\Lambda E {\bar\Lambda}{\bar E}+\Lambda E^c {\bar\Lambda}{\bar E^c})~~~.
\label{inv2}
\ee
The amplitudes of eq. (\ref{ae1}) are reproduced by
the combination ${\cal L}_e+8~\delta{\cal L}_e$.

On the other hand, there is no reason to prefer either the
result of eq. (\ref{ae1}) or that of eq. (\ref{ae2}).
The process $e^+e^-\to\grav\grav$ does not have a universal
low-energy behaviour. In the framework of non-linear
realizations, this freedom can be described by the invariant
Lagrangian ${\cal L}_e+\alpha~\delta{\cal L}_e$,
where $\alpha$ is a free parameter of the low-energy theory.

The process $e^+e^-\to\grav\grav\gamma$ suffers from a similar ambiguity
\cite{bfz4}.
Indeed the soft and collinear part of the cross-section, which is 
the dominant one, is associated to the initial-state radiation
and hence is determined by the cross-section for $e^+e^-\to\grav\grav$.
The total cross-section, with appropriate cuts on the photon 
energy and scattering angle, scales as $\alpha_{em} s^3/F^4$.
The photon energy and angular distributions
are not universal, as for the case of the goldstino angular distribution
in $e^+e^-\to\grav\grav$. They could be completely determined only through a 
computation performed in the fundamental theory.
From the non-observation of single-photon events above the SM background 
at LEP2, one can roughly estimate a lower bound on $\sqrt{F}$
of the order of the machine energy \cite{bfz4}. The precise value of this limit
would require the analysis of the relevant background, as well as
the inclusion of the above mentioned theoretical ambiguity. However,
in view of the quite strong power dependence of the cross-section
on the supersymmetry-breaking scale, one expects only a small
correction to the limit obtained by a rough dimensional estimate.
    
The partonic processes  $q{\bar q}\to\tilde{G}\tilde{G}\gamma$,
$q{\bar q}\to\tilde{G}\tilde{G}g$ are also expected to have cross-sections 
scaling as $\alpha_{em} s^3/F^4$ and  $\alpha_s s^3/F^4$ respectively.
The agreement between data and SM expectations in 
$p{\bar p}\to \gamma+{\slash E}_T+X$
and $p{\bar p}\to {\rm jet}+{\slash E}_T+X$ at the Tevatron collider could then be used to
infer a lower limit on $\sqrt{F}$. This is expected to be around
the typical total energy of the partonic subprocess, about $600$ GeV.
\section*{Acknowledgements}
\vskip 0.2cm
I would like to thank the organizers of the Ustron School for
their kind hospitality. Many thanks go to 
Andrea Brignole and Fabio Zwirner for the
pleasant collaboration on which this talk is based.
%

\end{document}